\documentclass[11pt]{article}
\usepackage{rotating}
  \usepackage{graphicx}
  \usepackage{psfrag}
  \usepackage{amsfonts}
\usepackage{amsmath,amssymb}
\usepackage{url}

\usepackage{natbib}
\usepackage{color}

\graphicspath{{figs/}}

\newtheorem{theorem}{Theorem}[section]
\newtheorem{lemma}[theorem]{Lemma}
\newtheorem{defn}[theorem]{Definition}

\newtheorem{question}[theorem]{Question}
\newtheorem{conjecture}[theorem]{Conjecture}

\newenvironment{example}[1][Example]{\begin{trivlist}
\item[\hskip \labelsep {\bfseries #1.}]}{\end{trivlist}}

\newcommand{\qed}{\nobreak \ifvmode \relax \else
      \ifdim\lastskip<1.5em \hskip-\lastskip
      \hskip1.5em plus0em minus0.5em \fi \nobreak
      \vrule height0.75em width0.5em depth0.25em\fi}

\newcommand{\R}{\mathrm{I\negthinspace R}}

\def\comment#1{\textit{[#1]}}
\def\comment#1{}

\begin{document}

\title{On the optimality of the neighbor-joining algorithm}
\author{Kord Eickmeyer\footnote{Informatics Institute, Humboldt University of Berlin, email: eickmeye@informatik.hu-berlin.de}
  \and Peter Huggins \footnote{Department of Mathematics, University of
    California at Berkeley, email: phuggins@math.berkeley.edu} \and Lior
  Pachter\footnote{Department of Mathematics, University of California
    at Berkeley, email: lpachter@math.berkeley.edu}\and Ruriko
  Yoshida \footnote{Department of Statistics, University of Kentucky, email: ruriko@ms.uky.edu}}
\date{}
\maketitle
\begin{abstract}

\end{abstract}
The popular neighbor-joining (NJ) algorithm used in phylogenetics
is a greedy algorithm for finding the
balanced minimum evolution (BME) tree associated to a dissimilarity
map. 
From this point of view, NJ is ``optimal'' when the
algorithm outputs the tree which minimizes the balanced minimum
evolution criterion. We use the fact that the NJ tree 
topology and the BME tree topology are determined by polyhedral
subdivisions of the spaces of dissimilarity maps ${\R}_{+}^{n \choose
  2}$ to study the optimality
of the neighbor-joining algorithm. In particular, we 
investigate and compare the polyhedral subdivisions for $n \leq 8$. 
A key requirement is the measurement of volumes of spherical polytopes
in high dimension, which we obtain using a combination of 
Monte Carlo methods and polyhedral algorithms.
We show that highly
  unrelated trees can be co-optimal in BME reconstruction, and that NJ
 regions are not convex. We obtain the $l_2$ radius for
  neighbor-joining for $n=5$ and we  conjecture that the ability of the
  neighbor-joining algorithm to 
 recover the BME tree  depends on the diameter of the BME tree.
\section{Introduction}

The popular neighbor-joining algorithm used for phylogenetic tree
reconstruction \cite{Saitou1987} has recently been ``revealed'' to be a
greedy algorithm for finding the balanced minimum evolution tree
associated to a
dissimilarity map \cite{Steel2006}. This means the following:

Let $D=\{d_{ij}\}_{i,j=1}^n$ be a 
dissimilarity map (this is an $n \times n$ symmetric matrix with zeroes
on the diagonals and non-negative real entries). 
The {\em balanced minimum
  evolution problem} is to find the tree $T$ that minimizes
\begin{equation}
\label{eq:bme}
\frac{1}{|o(T)|}\sum_{(x_1,\ldots,x_n) \in o(T)}\left[ \frac{1}{2}
    \sum_{i=1}^n d_{x_ix_{i+1}}\right]. 
\end{equation}Here $o(T)$ is the set of all cyclic permutations of the
leaves that arise from
    planar embeddings of $T$.
Denote by $p^T_{ij}$
the set of internal vertices in a tree $T$ on the path between $i$ and
$j$. Then (\ref{eq:bme})
is equivalent to minimizing 
\begin{equation}
\label{eq:bme_lp}
\sum_{ij} \lambda^T_{ij} d_{ij}
\end{equation}
where $\lambda^T_{ij} = \prod_{v \in p^T_{ij}} (deg(v)-1)^{-1}$ if $i \neq
j$ and $\lambda^T_{ii} = 0$.
This is an $NP$-hard linear programming problem \cite{Day1987} whose
relevance is given by the following theorem:
\begin{defn}
Let $T$ be a tree with $n$ leaves and $l:E(T) \rightarrow {\R}_{+}$ an
assignment of lengths to the edges. Then the length of $T$ with
respect to $l$ is
defined as
\[ l(T) = \sum_{e \in E(T)} l(e). \]
\end{defn}
\begin{theorem}[\cite{weightedsquares}]
Let $T$ be a binary tree with edge lengths given by $l:E(T)
\rightarrow {\R}_{+}$ and $D=\{d_{ij}\}_{i,j=1}^n$ a dissimilarity
map. If the variance of $d_{ij}$ is proportional to
$2^{|p^T_{ij}|}$ (i.e., $var(d_{ij}) = c 2^{|p^T_{ij}|}$ for some
constant
$c$) then (\ref{eq:bme_lp}) is the minimum variance tree length
estimator of $T$. Moreover, the weighted least squares
tree length estimate is equal to (\ref{eq:bme_lp}). 
\end{theorem}
This result provides a weighted least squares rationale for the
minimization of
(\ref{eq:bme_lp}), and highlights the importance of
understanding the {\em balanced minimum evolution polytope}:
\begin{defn}
The balanced minimum evolution polytope is the convex hull of the
vectors

\[
%\{[2^{-p_{12}^T},2^{-p_{13}^T},\ldots,2^{-p^T_{ij}},\ldots,2^{-p^T_{n-1
%      n}}] : T \mbox{ is a tree with } n \mbox{ leaves} \}. 
\{
\left[\lambda^T_{12},\lambda^T_{13},\ldots,\lambda^T_{ij},\ldots,\lambda^T_{n-1n}\right]
:
  T \mbox{ is a tree with }n \mbox{ leaves} \}
\]
\end{defn}

\begin{example}
There are four trees with $n=4$ leaves. They are the $3$ binary trees
and
the star-shaped tree. In this case the balanced minimum evolution
polytope is the
convex hull of the vectors:
\begin{eqnarray*}
& &
  \left[\frac{1}{4},\frac{1}{8},\frac{1}{8},\frac{1}{8},\frac{1}{8},\frac{1}{4}\right]
  \quad \mbox{T is the tree with leaves 1,2 separated from 3,4,}\\
& &
  \left[\frac{1}{8},\frac{1}{4},\frac{1}{8},\frac{1}{8},\frac{1}{4},\frac{1}{8}\right]
 \quad   \mbox{T is the tree with leaves 1,3 separated from 2,4,}\\
& &
  \left[\frac{1}{8},\frac{1}{8},\frac{1}{4},\frac{1}{4},\frac{1}{8},\frac{1}{8}\right]
    \quad \mbox{T is the tree with leaves 1,4 separated from 2,3,}\\
& &
  \left[\frac{1}{3},\frac{1}{3},\frac{1}{3},\frac{1}{3},\frac{1}{3},\frac{1}{3}\right]
    \quad \mbox{T is the star-shaped tree.}\\
\end{eqnarray*}
The balanced minimum evolution polytope in this case is a triangle in
$\R^6$. Note that the star-shaped tree is in the interior of the
triangle.
\end{example}
For any dissimilarity map, the trees which minimize (\ref{eq:bme_lp})
will be vertices of the balanced minimum evolution polytope; these are
always the binary trees. In fact, for such trees $\lambda^T_{ij} =
2^{-|p^T_{ij}|}$ (this is Pauplin's formula \cite{circularorderings}).
The BME polytope lies in $\R^{{n \choose 2}}$ and has dimension ${n
  \choose 2} - n$. The normal fan \cite{ascb} of the BME polytope gives rise to {\em
  BME
  cones} which form a polyhedral subdivision of the space of
  dissimilarity maps $\R_{+}^{{n
    \choose 2}}$. They describe, for each tree $T$, those dissimilarity
maps for which $T$ minimizes (\ref{eq:bme_lp}).
We discuss the polytope in more detail in Section 2.

The neighbor-joining algorithm is a greedy algorithm for finding an
approximate solution to (\ref{eq:bme_lp}). We omit a
detailed description of the algorithm here (readers can consult
\cite{Steel2006}), but we do mention the crucial fact
that the selection criterion is linear in the dissimilarity map
\cite{Bryant2005}. Thus, 
the NJ algorithm will pick 
pairs of leaves to merge in a particular order and output a particular
tree $T$ if and
only 
if the
pairwise distances satisfy a system of linear inequalities, whose
solution
set forms a polyhedral cone in  ${\R}^{n \choose 2}$.  We call such a
cone a 
{\em neighbor-joining cone,} or {\em NJ cone}.  Thus the NJ algorithm
will
output a particular tree $T$ if and only if the distance data lies in a union of NJ
cones.
In Section 3 we show that the NJ cones partition  ${\R}^{n \choose
  2}$, but do not form a fan.  This has
important implications for the behavior of the NJ algorithm. 

Our main result is a comparison of the neighbor-joining cones with the
normal fan of the balanced minimum evolution polytope. This means that
we
characterize those dissimilarity maps for which neighbor-joining,
despite being a greedy algorithm, is able to identify the balanced minimum
evolution tree. These results are discussed in Section 4. 

\section{The balanced minimum evolution polytope}
Throughout this paper, let $\{1, \cdots, n\}$ be the set of
taxa.  
Recall there are $2n - 3$ edges in an unrooted tree with $n$ leaves.
For a fixed tree topology $T$, let $B_{T}$ be the ${n \choose 2} \times
(2n-3)$
matrix with rows indexed by pairs of leaves and columns indexed by edges
in $T$ defined as follows:
\[
B_{T}(\{a,b\}, e)  =  \begin{cases}
1 & \text{if edge } e \text{ is in the path from leaf }a \text{ to leaf
} b,\\
0   & \text{otherwise}.
\end{cases}
\]
For example, for the tree in Figure \ref{fig:fivetree},
\[
B_T \qquad = \qquad 
\left(\begin{array}{ccccccc}
1 & 1& 0 & 0 & 0 & 0 & 0\\
1 & 0 & 1 & 0 & 0 & 1 & 0 \\
0 & 1 & 1 & 0 & 0 & 1 & 0\\
1 & 0 & 0 & 1 & 0 & 1 & 1\\
0 & 1 & 0 & 1 & 0 & 1 & 1\\
0 & 0 & 1 & 1 & 0 & 0 & 1\\
1 & 0 & 0 &  0 & 1 & 1 & 1\\
0 & 1 & 0 & 0 &  1 & 1 & 1\\
0 & 0 & 1 & 0 & 1 & 0 & 1\\
0 & 0 & 0 & 1 & 1 & 0 & 0 \\
\end{array}\right).
\]
Given edge lengths $l:E(T) \rightarrow \R_{+}$ we let $\mathbf{b}$ be
the vector with components $l(e)$ as $e$ ranges over $E(T)$.
Any dissimilarity map $\mathbf{d}$ (encoded as a row vector) can now be
written as
\[
\mathbf{d} = B_{T} \mathbf{b} + \mathbf{e}.
\]
where $\mathbf{e}$ is a vector of ``error'' terms that are zero when
$\mathbf{d}$ is a tree metric. 

The weighted least squares solution for the edge lengths $\mathbf{b}$
assuming a variance matrix $V$ with diagonal entries $v_{ij} =
\lambda^T_{ij}$ (as defined in the introduction) and 
dissimilarity map $\mathbf{d}$ is given by

\[
\hat{\mathbf{b}} = (B_{T}^{t}V^{-1}B_{T})^{-1}B_{T}^{t}V^{-1}\mathbf{d},
\]
where ${\cdot}^{t}$ denotes
matrix transpose. The length of $T$ with respect to the least squares
edge lengths is then
\[
l(T) = \mathbf{v}_{T} \cdot \mathbf{d},
\]
 where $\mathbf{v}_{T} = V^{-1}B_{T}(B_{T}^{t}V^{-1}B_{T})^{-1}{\bf 1}$ and ${\bf 1}$ is the 
vector of all 1's.  We call the vectors $\mathbf{v}_T$ the balanced minimum
evolution vectors (or BME vectors). In the case of Figure
\ref{fig:fivetree}, the BME vector is 
\[
\mathbf{v}_T =
\left[\frac{1}{2},\frac{1}{4},\frac{1}{8},\frac{1}{8},\frac{1}{4},\frac{1}{8},\frac{1}{8},\frac{1}{4},\frac{1}{4},\frac{1}{2}\right].
\]

The BME method
is equivalent to minimizing the linear functional $\mathbf{v}_T \cdot \mathbf{d}$
over
all BME vectors for all tree topologies $T$.  The BME polytope is the
  convex hull of all BME vectors in $\R^{n \choose 2}$. The following
  facts follow from the definition of the balanced minimum
  evolution
  tree:
\begin{lemma}
The vertices of the BME polytope are the BME
 vectors of binary trees.  The BME vector of
  the star phylogeny lies in the interior of the BME polytope, and all
 other BME vectors  lie on the boundary of the BME polytope.
\end{lemma}

  The normal fan \cite{ascb} of a BME polytope partitions the space
  $\R^{n
  \choose 2}$ of
  dissimilarity maps into cones, one for each tree.
  We call these {\em BME cones}.  They completely characterize the BME
  method:  $T$ is the
  BME tree topology if and only if the dissimilarity map $D$ lies in the BME cone
  of
  $T$.

  For a node $a$ its \emph{shift
  vector} $\mathbf{s}_a$ is the dissimilarity map in which $a$ 
  is distance 1 from all other leaves, and all other distances are 0.
  According to 
  \cite{circularorderings}, for a tree $T$, $(\mathbf{v}_T)_{ab}$
gives the
  probability that $a$ will immediately precede $b$ in a random circular
  ordering of $T$.
  Thus the dot-product of a BME vector with a shift vector
  must necessarily equal 1, and 
  in fact the lineality space of BME cones is spanned by shift vectors.
  So when we describe a BME cone we will always
  describe just the pointed component, i.e. modulo the lineality space
  of shift vectors.

\begin{table}
\label{table:fvec}
\begin{center}
\begin{tabular}{|l|l|l|}
\hline
\#leaves & dim(BME polytope) & $f$-vector \\ 
\hline
4      &  2  & (3,3)   \\
\hline
5      &  5  & (15, 105, 250, 210, 52)   \\
\hline 
6      & 9   & (105, 5460, ?, ?, ?, 90262)   \\ 
\hline 
7      & 14  & (945, 445410, ?, ?, ?, ?, ?)   \\   \hline
\vdots & \vdots & \vdots \\ \hline
$n$ & ${n \choose 2}-n$ & (($2n-5$)!!,?,\ldots) \\\hline

\end{tabular}
\end{center}
\caption{The $f$-vector for small BME polytopes.}
  \end{table}

  As part of our computational study, we computed the BME polytope and
  BME
  cones for trees with $n=4,5,6,7,8$ leaves using
  the software {\tt polymake} \citep{Gawrilow2000}. In Table 1 we display some of the components of $f$-vectors we
  were able to compute. This provides information about the polytopes:
  the $i$th component of an $f$-vector of a  
  polytope is the number of faces of dimension $i-1$. For example, the
  first component in each vector in Table 1 is the number
  of $0$-dimensional faces (vertices) of the
  corresponding BME polytope, i.e., the number of binary trees.

\begin{figure}[!ht]
\begin{center}
 \includegraphics[scale=0.7]{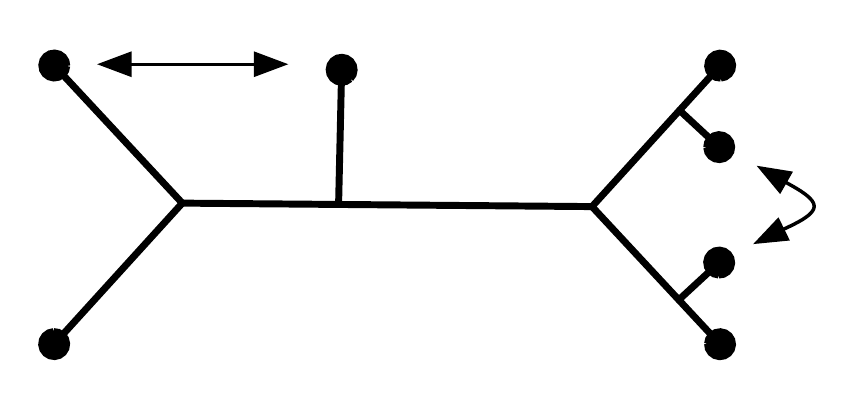}
\end{center}
\caption{The non-edges on the BME polytope for $n = 7$.  Two trees will
  form a non-edge if and only if they 
         are 3-cherry trees that differ by the pair of leaf exchanges
  shown in the figure. There are two ways to perform each leaf-exchange,
         so each binary tree with three cherries is not adjacent to
  4 trees.}
\label{fig:nonedge}
\end{figure}

  We found that the edge graph of the BME polytope is the complete graph
  for $n=4,5,6$ which means that for every pair of trees
$T_1$ and $T_2$ with the same number ($\leq 6$) of leaves, there is
  a  dissimilarity map for which $T_1$ and $T_2$ are (the only)
  co-optimal BME trees. However, for $n=7$, the BME polytope does in
  fact
  have one combinatorial type of non-edge.  Namely,
  two bifurcating trees with seven leaves and three cherries (two leaves
  adjacent to the same node in the tree) will form a non-edge if and
  only if they
  are related by
  two leaf exchanges as depicted in Figure \ref{fig:nonedge}. This
  completely characterizes the non-edges for $n=7$.  It is an
  interesting open problem to characterize the
non-edges of the BME polytope in general.

\section{Neighbor-joining cones}
\label{sec:nj}

The neighbor-joining algorithm takes as input a dissimilarity map and
outputs a tree. The tree is constructed ``one cherry at a time''. This
means that at each step leaves $a,b$ are picked to be a cherry by 
minimizing the \emph{Q-criterion}. The Q-criterion is given by the
formula
\begin{equation}
\label{eqn:qcrit}
q_{ab} := (n-2)d_{ab} - \sum_{k=1}^{n} d_{ak} - \sum_{k=1}^{n}
d_{kb}.
\end{equation}

The nodes $a,b$ are replaced by a single node $z$, and new
distances $d_{zk}$ 
are obtained by a straightforward linear combination of the
original pairwise distances:  $d_{zk} := \frac{1}{2}(d_{ak} + d_{bk} - 
d_{ab})$.  Then the NJ method is applied recursively.  
% this was 2d_{ab} in an earlier version, but it should be d_{ab} --
%Kord

We note that since new distances $d_{zk}$ are always linear
combinations of the previous distances, all Q-criteria computed
throughout the NJ algorithm are linear combinations of the original
pairwise distances.  Thus, for a fixed $n$, for every possible
ordering $\sigma$ of picked cherries that results in one of the trees
$T$ with $n$ leaves there is a polyhedral cone $C_{\sigma} \subset
\R^{n \choose 2}$ of dissimilarity maps. The set of all
neighbor-joining cones is denoted by ${\cal C}_n$. Their union
$\bigcup_{C \in {\cal C}_n} C$ is all of of $\R^{n \choose 2}$, and
the intersection of any two cones is a subset---but not necessarily a
face---of the boundary of each of the cones. Given an input from the
interior of $C_{\sigma}$, the NJ algorithm will pick the cherries in
the order $\sigma$ and output the corresponding tree. For inputs
$\mathbf{d}$ on the boundary of one (and therefore at least two) of
the cones, the order in which NJ picks cherries is undefined, because
at some point there will be two cherries both of which have minimal
Q-criterion.
%The NJ algorithm will pick
%cherries in the order $\sigma$ iff 
%the input lies in the cone $C_{\sigma}$.
We call the cones
$C_{\sigma}$ {\em neighbor-joining cones}, or {\em NJ cones}. 

\begin{figure}[ht]
\begin{center}
\includegraphics[scale=0.7]{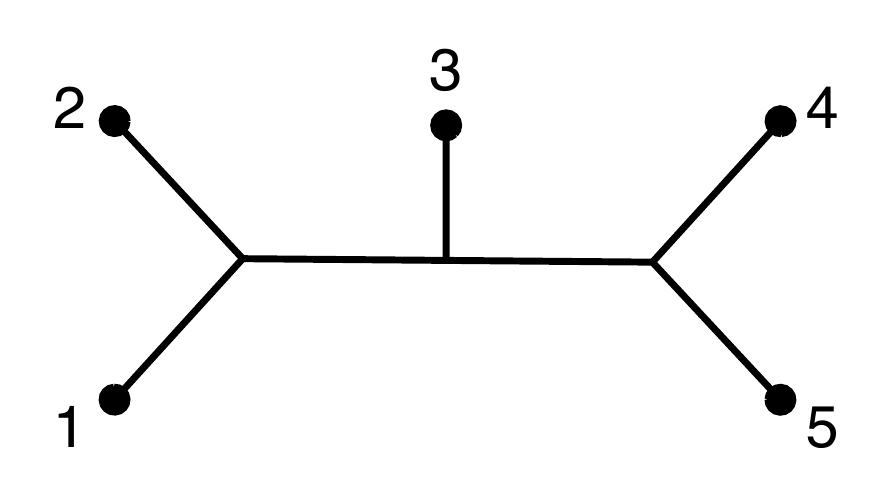}
%\resizebox{0.6\textwidth}{!}{\input fivetree.pstex_t}
\end{center}
\caption{A tree with five leaves.}
%\caption{(a) A tree with five taxa (b) The same tree with all edges
%adjacent to leaves reduced to length zero. The remaining two edges
%have lengths $\alpha$ and $\beta$.}
\label{fig:fivetree}
\end{figure}

\begin{example}
There is only one unlabeled binary tree with $5$ leaves and there are 15 distinct labeled
trees. For each labeled tree, there are two ways in
which a cherry might be picked by the NJ algorithm in the first
step. For instance, neighbor-joining applied to any dissimilarity map in
$C_{12,45}$ or $C_{45,12}$ will produce the tree in Figure
\ref{fig:fivetree}.  There are a total of $30$ NJ cones for $n = 5$.
\end{example}

We note that all Q-criteria for shift vectors equal $-2$, so adding any
linear combination of
shift vectors to a dissimilarity map does not change the relative values
of the Q-criteria. Also, after picking a cherry,
the reduced distance matrix of a shift vector is again a shift
vector. Thus,
for any input vector $\mathbf{d}$, the behavior of the NJ algorithm on
$\mathbf{d}$ will be the same as on $\mathbf{d} + \mathbf{s}$ if
$\mathbf{s}$ is any linear combination of shift vectors. 
In fact it can be shown that the lineality space of NJ cones is spanned
by shift vectors, just as for BME cones.  
So from now on, when we refer to NJ cones,
we will mean the pointed portion of the cone, i.e. modulo the lineality
space.

\begin{theorem}
\label{thm:normalfantheorem}
The cones in ${\cal C}_n$ are not the normal fan of any polytope for $n
\geq 5$. 
\end{theorem}
To prove this theorem it is necessary to understand the geometry of the
NJ cones. We describe the case $n=5$ in detail; it also suffices to
prove the theorem.

\begin{table}[!h]
\label{table:rays}
\begin{center}
\begin{tabular}{|c|c|c|}
\hline Type & rays & Cones\\\hline
I & \begin{tabular}{c}
  $(-3, 5, -3, -1, 5, -3, -1, 1, 1, -1)$\\
  $(-3, 5, -3, -1, 1, 1, -1, 5, -3, -1)$\\
  $(5, -3, -3, -1, -3, 5, -1, 1, 1, -1)$\\
  $(1, 1, -3, -1, -3, 5, -1, 5, -3, -1)$\\
  $(5, -3, -3, -1, 1, 1, -1, -3, 5, -1)$\\
  $(1, 1, -3, -1, 5, -3, -1, -3, 5, -1)$\end{tabular} &
  \begin{tabular}{c}
    $C_{23,45},C_{23,15},C_{23,14},C_{12,34}^*,C_{34,12}^*$\\
    $C_{23,45},C_{23,15},C_{23,14},C_{12,35}^*,C_{35,12}^*$\\
    $C_{23,45},C_{23,15},C_{23,14},C_{24,13}^*,C_{13,24}^*$\\
    $C_{23,45},C_{23,15},C_{23,14},C_{25,13}^*,C_{25,13}^*$\\
    $C_{23,45},C_{23,15},C_{23,14},C_{24,35}^*,C_{35,24}^*$\\
    $C_{23,45},C_{23,15},C_{23,14},C_{25,34}^*,C_{25,34}^*$
   \end{tabular} \\\hline
II & \begin{tabular}{c}
$(-1, 1, -1, 1, 1, -1, -1, 1, 1, -1)$\\\\[1.3ex]
$(-1, 1, -1, -1, 1, 1, 1, 1, -1, -1)$\\\\[1.3ex]
$(1, 1, -1, -1, -1, 1, -1, 1, -1, 1)$\\\\[1.3ex]
$(1, -1, -1, 1, -1, 1, -1, 1, 1, -1)$\\
\end{tabular} & \begin{tabular}{c}
$C_{12,45}, C_{12,34}, C_{23,45}, C_{23,15}, C_{34,15},$\\
$C_{34,12}, C_{45,23}, C_{45,12}, C_{15,34}, C_{15,23}$\\[1.3ex]

$C_{12,45}, C_{12,35}, C_{23,45}, C_{23,14}, C_{35,14},$\\
$C_{35,12}, C_{45,23}, C_{45,12}, C_{14,35}, C_{14,23}$\\[1.3ex]

$C_{25,14}, C_{25,13}, C_{23,14}, C_{23,45}, C_{13,45},$\\
$C_{13,25}, C_{14,23}, C_{14,25}, C_{45,13}, C_{45,23}$\\[1.3ex]

$C_{24,15}, C_{24,13}, C_{23,15}, C_{23,45}, C_{13,45},$\\
$C_{13,24}, C_{15,23}, C_{15,24}, C_{45,13}, C_{45,23}$
\end{tabular}
\\\hline
III & \begin{tabular}{c}
$(1, -1, -1, 1, 1, -1, -1, -1, 3, -1)$\\\\[1.3ex]
$(1, -1, -1, -1, -1, 3, 1, 1, -1, -1)$\\\\[1.3ex]
$(1, -1, -1, 1, 1, -1, -1, -1, 3, -1)$\\\\[1.3ex]
$(1, -1, -1, -1, -1, 3, 1, 1, -1, -1)$\\
\end{tabular} & \begin{tabular}{c}
$C_{23,45}, C_{23,15}, C_{12,45}, C_{12,35}, C_{24,15}, C_{24,35},$\\
$C_{35,24}, C_{35,12}, C_{15,24}, C_{15,23}, C_{45,12},
C_{45,23}$\\[1.3ex]
 
$C_{23,45}, C_{23,14}, C_{12,45}, C_{12,34}, C_{25,14}, C_{25,34},$\\
$C_{34,25}, C_{34,12}, C_{14,25}, C_{14,23}, C_{45,12},
C_{45,23}$\\[1.3ex]
 
$C_{23,45}, C_{23,15}, C_{13,45}, C_{13,25}, C_{34,15}, C_{34,25},$\\
$C_{25,34}, C_{25,13}, C_{15,34}, C_{15,23}, C_{45,13},
C_{45,23}$\\[1.3ex]
 
$C_{23,45}, C_{23,14}, C_{13,45}, C_{13,24}, C_{35,14}, C_{35,24},$\\
$C_{24,35}, C_{24,13}, C_{14,35}, C_{14,23}, C_{45,13}, C_{45,23}$ 
\end{tabular}\\\hline
\end{tabular}
\end{center}
\caption{The 14 rays of the cone $C_{23,45}$. Each ray is determined
  by a vector shown in the second column. The third column shows, for
  each ray, which cones it belongs to. If a cone is starred then the
  ray is inside the cone, but not a ray of it.}
\end{table}

We begin by noting that all of the NJ cones are equivalent under the
action of the symmetric group on five elements ($S_5$), where an
element of $S_5$ permutes the five taxa or, equivalently, the rows and
columns of the input distance matrix. Each NJ cone is defined by $({5
  \choose 2} - 1) + ({4 \choose 2} - 1) = 14$ inequalities that are
implied by the Q-criteria as the NJ algorithm picks the two
cherries. The cones are $5$-dimensional, and their intersection with a
suitable hyperplane leaves a four dimensional polytope $P$. The
$f$-vector of $P$ is $(14,32,27,9)$.

The $30$ cones share many of their rays, giving a total of $82$
rays which decompose into three
orbits under the action of $S_5$. We refer to the types of rays as
Type I, Type II and Type III. Each cone has $6$ rays of type I, $4$
rays of type II and $4$ rays of type III. Each ray of type I is the
common ray of $3$ cones, and belongs to $2$ other cones of which it
is not a ray (i.e. it is in the interior of a face). Note that this
implies that the cones cannot form a fan. The type II rays are
contained in $10$ cones each, and the type III rays in $12$. Type II
and III rays are rays of all cones which contain them. For the cone
$C_{23,45}$, this information is tabulated in Table 2.

%We begin by describing the
%combinatorics of the $14$ rays of each of the NJ cones (this suffices
%to prove the theorem), and then proceed to describe the remaining
%facet structure of $P$.

%A computation using
%\texttt{polymake} \citep{Gawrilow2000} reveals that two inequalities 
%$\mathbf{h}_{9,1}$
%and $\mathbf{h}_{9,2}$ do not define facets.
%There are $82$ rays altogether, and they decompose into three orbits
%under the 
%action of $S_5$. We refer to the types of rays as Type I, Type II and
%Type III. Each cone contains 6 rays of Type I, 4 of Type II and 4 of
%Type III. There are a total of 60 rays of Type I, 12 of Type II and 10
%of Type III. This information is tabulated in Table \ref{table:rays}

We note that the rays of NJ cones are minimal intersections of NJ cones,
and thus give dissimilarity maps for which the NJ
algorithm is least stable. 
% Note that the ray ??? is contained in the
%cones $C_{23,45},C_{23,15}$ and $C_{23,14}$ but not in $C_{12,34}$ and
%$C_{34,12}$ which also contain it. It follows that ${\cal C}_n$ is not
%a
%fan. \qed
\begin{example}
Consider two alignments of 5 sequences that are to be used to construct
a tree. These may consist of two different genes and for each of them
the homologs among 5 genomes. 
Suppose that distances are estimated using the Jukes--Cantor
correction \cite{Jukes1969,ascb} separately for each set of sequences. That is, for the first
set of sequences
\[
(D_1)_{ij} = -\frac{3}{4}\log(1- \frac{4}{3} f_{ij})\]
where $f_{ij}$ is the fraction of different nucleotides between
sequences $i$ and $j$ in the first set and for the second set
\[
(D_2)_{ij} = -\frac{3}{4}\log(1- \frac{4}{3} g_{ij})
\]
where $g_{ij}$ is the fraction of different nucleotides between
sequences $i$ and $j$ in the second set. 

If the fractions $f_{ij}$ and $g_{ij}$ are given by
\[
f := \left(\begin{array}{ccccc}
0 &0.054187 &0.151108 &0.368136 &0.054198 \\
0.054187 &0 &0.151117& 0.054198& 0.36813 \\
0.151108& 0.151117 &0 &0.054187& 0.054198 \\
0.368136& 0.054198 &0.054187& 0& 0.151108 \\
0.054198 &0.36813 &0.054198 &0.151108& 0 \\
\end{array}\right)\text{ and }
\]
\[
g : = \left(\begin{array}{ccccc}
0 & 0.151068  &0.05414  &0.368161  &0.104517 \\
0.151068 & 0  &0.054245 & 0.054245 & 0.395699 \\
0.05414  &0.054245 & 0 & 0.151068 & 0.194428 \\
0.368161 & 0.054245  &0.151068  &0 & 0.104421 \\
0.104517  &0.395699 & 0.194428 & 0.104421  &0\\
\end{array} \right)\,
\]
then we obtain 
\[
D_1 = \left(\begin{array}{ccccc}
0 & 0.056244  &0.168744 & 0.506257 & 0.056256 \\
0.056244  &0  &0.168755 & 0.056256  &0.506245 \\
0.168744 & 0.168755  &0 & 0.056244  &0.056256 \\
0.506257  &0.056256 & 0.056244 & 0 & 0.168744 \\
0.056256 & 0.506245 & 0.056256 & 0.168744  &0 \\
\end{array}\right) \text{ and }
\]
\[
D_2 = \left(\begin{array}{ccccc}
0 & 0.168694 & 0.056194  &0.506306 & 0.112556 \\
0.168694  &0 & 0.056307  &0.056307  &0.562445 \\
0.056194 & 0.056307  &0  &0.168694  &0.225056 \\
0.506306 & 0.056307  &0.168694 & 0  &0.112444 \\
0.112556  &0.562445  &0.225056 & 0.112444 & 0 \\
\end{array}\right) \, .
\]
Notice that the vector representation of $D_1$ lies in the cone 
$C_{12, 45}$ and the vector representation of $D_2$ lies in the cone
$C_{45, 12}$.  Thus NJ returns the same tree topology for both
$D_1$ and $D_2$.

If we concatenate the alignments and combine the data to build one
tree, then we estimate the distances using the average of $f$ and $g$:
\[
\frac{1}{2}\left(f + g \right)= \left(\begin{array}{ccccc}
0 & 0.102628 & 0.102624 & 0.368148&  0.079357 \\
0.102628&  0 & 0.102681 & 0.054222&  0.381915 \\
0.102624 & 0.102681 & 0&  0.102628 & 0.124313 \\
0.368148&  0.054222&  0.102628&  0&  0.127765 \\
0.079357 & 0.381915&  0.124313 & 0.127765&  0 \\
\end{array}\right) \, .
\]
Using this frequency matrix we obtain the distance matrix $D_3$ via the
Jukes--Cantor correction:
\[
D_3 = \left(\begin{array}{ccccc}
0 & 0.110364&  0.110359&  0.506281 & 0.083878 \\
0.110364 & 0 & 0.110425 & 0.056281&  0.533818 \\
0.110359 & 0.110425 & 0&  0.110364&  0.135917 \\
0.506281 & 0.056281&  0.110364&  0 & 0.140066 \\
0.083878 & 0.533818 & 0.135917&  0.140066&  0 \\
\end{array}\right) \, .
\]

However, the vector representation of $D_3$ lies in
the cone $C_{24,15}$, which means that neighbor-joining returns a
different tree topology
for $D_3$. This example provides a distance-based reconstruction analog
to the
recent mixture model results of \cite{Matsen2007}.
\end{example}

\begin{figure}[!ht]
\label{fig:facets}
\begin{center}
\includegraphics[scale=0.6]{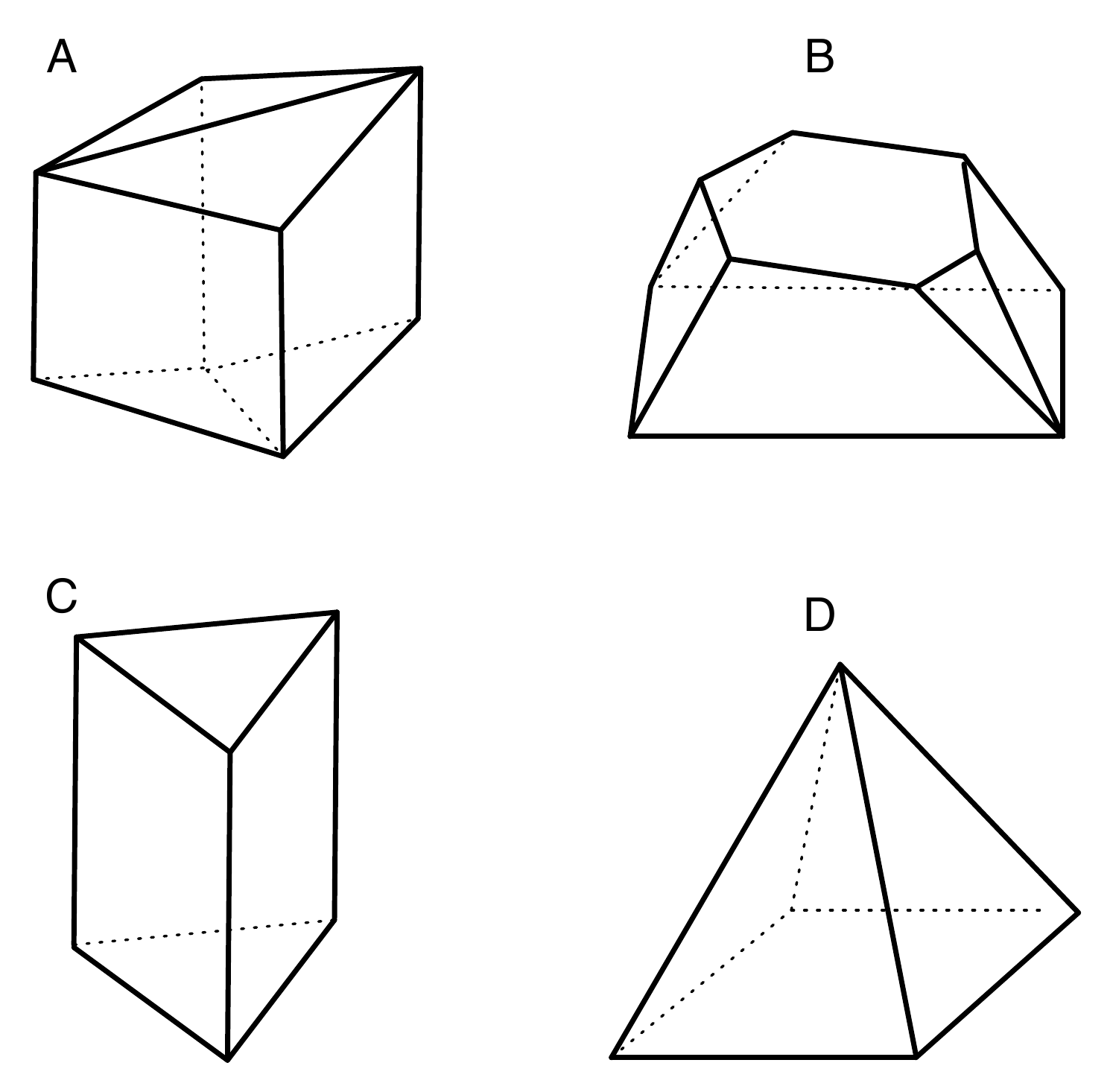}
\end{center}
\caption{The four types of facets of $P$.}
\end{figure}

An analysis of the rays of ${\cal C}_n$ suffices to prove Theorem
\ref{thm:normalfantheorem} but the facet structure of each cone is also
informative, and we were able to obtain complete information for
$n=5$. 
The types of facets constituting each cone are shown in Figure
\ref{fig:facets}. Each cone consists of one Type A facet, two Type B
facets, two Type C facets and four Type D facets. These facets intersect
as follows: Type A facets are shared 
by pairs of cones of the form $C_{ab,cd},C_{cd,ab}$. Type B 
facets are shared by pairs of cones of the form $C_{ab,de},C_{ab,ce}$; there are two such pairs for each cone. Two of the square
facets of a Type A facet belong to Type B facets, and a pair of Type B
facets share a hexagon consisting of six Type I rays. The remaining
two square facets of a Type A facet form Type C facets with two Type
I rays. The four triangular facets of a Type A facet form Type D facets
(Egyptian pyramids) with two Type I rays. 

We used our description of the NJ cones to examine the $l_2$ distance
between tree metrics and the boundaries of NJ cones. Without loss of
generality, by shifting the leaves that in the cherries, we can assume the tree metric is of the form
\[
D_T \qquad = \qquad 
\left(\begin{array}{ccccc}
0 & 0 & \alpha & \alpha+\beta & \alpha+\beta\\
0 & 0 & \alpha & \alpha+\beta & \alpha+\beta\\
\alpha & \alpha & 0 & \beta & \beta\\
\alpha+\beta & \alpha+\beta & \beta & 0 & 0\\
\alpha+\beta & \alpha+\beta & \beta & 0 & 0\\
\end{array}\right)
\]
where $\alpha$ and $\beta$ are the internal branch lengths, $\alpha \geq
\frac{1}{2}$ and $\alpha+\beta=1$. It is easy to see that $D_T \in
C_{12,45}$ confirming the consistency of neighbor-joining. The cone
$C_{12,45}$ contains $9$ faces, but we may ignore one of them
($C_{45,12}$) as it corresponds to the same tree. The distance to the
closest of the remaining eight faces is
\[  d(D_T,(C_{12,45} \cup C_{45,12})^{c}) =
\frac{1-\alpha}{\sqrt{3}}. \]
We summarize this as follows:
\begin{theorem}
The $l_2$ radius of neighbor-joining for $5$ taxa is $\frac{1}{\sqrt{3}}
\approx 0.5773$.
\end{theorem}
This is slightly
  larger than the $l_{\infty}$ radius of $\frac{1}{2}$ given by
  Atteson's theorem \cite{Atteson99}. It is an interesting problem to
  compute the $l_2$
  radius for neighbor-joining with more taxa.

The description of the NJ cones we have provided can also be used in
practice to evaluate the robustness of the algorithm when used with a
specific dataset.
For $n = 5$, we examined data simulated from subtrees of the two tree 
models $T_1$ and $T_2$ in \citep{Ota2000} with the Jukes-Cantor model
and the Kimura 2-parameter models \citep{ascb}.
For each of $40,000$ simulations, we calculated the $\ell_2$-distance
between the NJ cone of the given tree
and the maximum likelihood estimates for the pairwise distances (see
supplementary material). These
show that in many cases the maximum likelihood estimates lie very close
to the boundary. In such cases, one must conclude that the NJ tree is
possibly incorrect due to the variance in the distance estimates.

\section{Optimality of the neighbor-joining algorithm}

In order to study the optimality of the neighbor-joining algorithm, we
compared the BME cones with the NJ cones. Such a comparison involves
intersecting the cones with the $({n \choose 2}-1)$-sphere (in the first
orthant) and
then studying the volumes of their intersection by computing the
standard Euclidean volume of the resulting surfaces. 
These surfaces are an intersection of closed hemispheres,
i.e. {\em spherical polytopes}.  
Computing Euclidean volumes of (non-spherical) polytopes is a standard
problem that is usually solved by triangulating
and summing the volumes of the simplices.
However there has been no publicly available software
developed for
computing or approximating 
volumes of spherical polytopes of dimension $> 3$ using this method.
One possible reason for this is that in higher dimensions the volumes of 
spherical simplices are given by complicated analytical formulas
\citep{sphericalsimplex}
whose computational complexities are unknown.

We implemented two approaches in MATLAB (using {\tt polymake} as a
preprocessing step) for approximating the volume of a spherical polytope
$P$.  
One approach is  trivial:  it simply
samples uniformly from the sphere, and counts how many points are inside
$P$.  This
approach is particularly suitable if $P$ has large volume, or if many
spherical polytopes are being simultaneously measured
which partition the sphere, as is the case for NJ and BME cones.  The
second approach is suitable
for spherical polytopes having small volume. We used this approach for
computing the volumes of {\em consistency cones} \cite{Mihaescu2007}
which we discuss briefly in the Discussion section.

The second approach begins by computing a triangulation of the vertices
of $P$ 
with additional interior points of $P$ added.  This triangulation
defines a simplicial mesh $M$
which is obtained by replacing each
spherical simplex with the corresponding Euclidean simplex having the
same vertices.  The volume of $M$ (i.e. the
sum of the volumes of the simplices in the mesh) is already an
approximation to the volume of $P$.  We refine this estimate
by Monte Carlo estimation of the average value of the Jacobian from
$M$ to $P$.  
This requires sampling uniformly from $M$, which is straightforward 
and can be done very quickly in $O(m + kd + k \log k)$ time, where $m$
is the number of simplices in the 
mesh, $k$ is the number of samples, and $d$ is the dimension.  

Our main results on the the optimality of NJ for $n = 5,6,7,8$ taxa are
summarized in Table 3. Each row of the table describes
one type of tree. Trees are classified by their topology. A $k$-cherry
tree is a tree with $k$ cherries. The NJ volume column shows the volume
of that
part of the
positive orthant of dissimilarity maps for which the NJ tree is of the
specified type. Similarly, the BME volume column shows the same
statistic
for BME trees. Finally, NJ accuracy shows the fraction of the BME cone
that overlaps the NJ cone. In other words, NJ accuracy is a measure of
how
frequently NJ will find the BME tree for a dissimilarity map that is
chosen at random.

\begin{table}
\label{table:main}
\begin{center}
\begin{tabular}{|c|c|r|r|r|r|}
\hline
\#taxa & tree shape & \#trees & NJ vol & BME vol & NJ accuracy \\
\hline
4 & unique & 3 & 100\% & 100\% & 100\%\\ \hline
5      &     unique    & 15     & 100\%      & 100\%    & 98.06\%
\\
\hline
6      & 3-cherry   & 15     & 18.50\%      & 18.57\%     & 90.39\%
\\
\hline
6      & caterpillar & 90     & 81.54\%      & 81.45\%     & 91.33\%
\\
\hline
7      & 3-cherry    & 315    &  45.32\%    &   44.58\%    &  82.42\%
\\
\hline
7      & caterpillar  &  630  &    54.68\%    &   55.42\%   & 78.85\%
\\
\hline
8      & 4-cherry     &   315   &   6.48\%    &  6.36\%  & 70.12\%
\\
\hline
8      & 3-cherry     &   2520   &  27.12\%   &   25.84\%   & 69.93\%
\\
       & (two are neighbors) & &         &          &
\\
\hline
8      & 3-cherry    &   2520   &   35.67\%    &  34.54\%   & 71.63\%
\\
       & (none are neighbors)& &       &         &
 \\
\hline
8      & caterpillar &   5040    &  30.73\%    &  33.24\%   &  61.75\%
\\
\hline
\end{tabular}
\end{center}
\caption{Comparison of NJ and BME cones.}
\end{table}

\begin{figure}[ht]
\label{fig:mistakes}
\begin{center}
\includegraphics[scale=0.7]{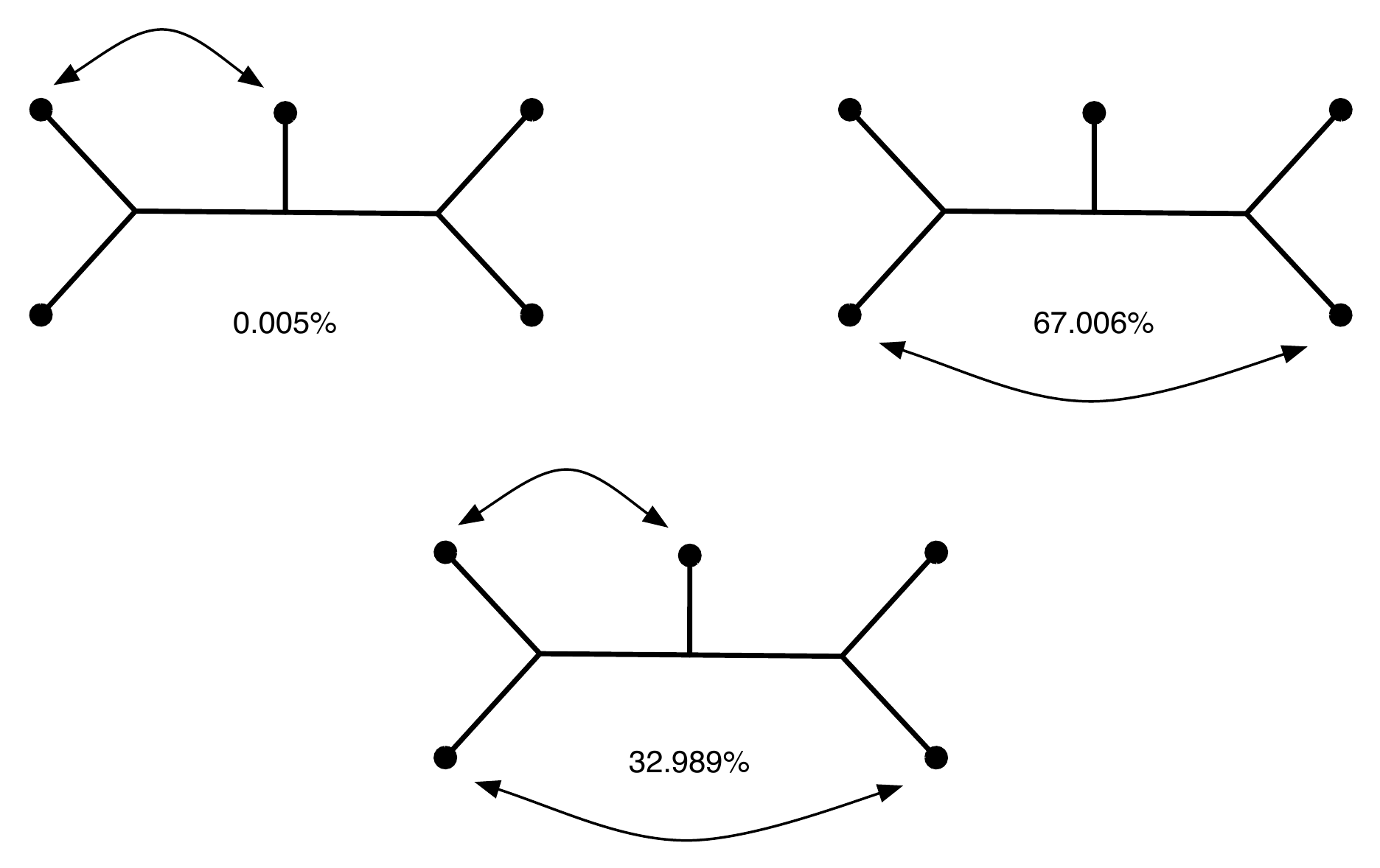}
\end{center}
\caption{Frequencies of the three possible types of NJ trees that may
  picked instead of the BME tree for $n=5$ leaves. Neighbor-joining
  agrees with the BME tree 98.06\% of the time.}
\end{figure}

We also classified and measured the intersections of NJ and BME cones in
which the NJ tree differs from the BME tree.
Many of these intersection cones are equivalent under the action of
$S_n$ on the leaf labels, particularly as the stabilizer of the BME tree
permutes the leaf labels in the NJ tree.  In fact, for $n = 5$ taxa
there are only three types of mistakes
that the NJ algorithm can make when it fails to reproduce the BME tree.
These are depicted in Figure \ref{fig:mistakes} and the 
normalized spherical volumes of corresponding NJ/BME intersection cones
are given.

Figure 4 can be interpreted as follows: For a random dissimilarity map,
if the NJ algorithm does not produce the BME tree,
then with probability $0.67$ it produces the tree on the right, and if
not then it almost always produces the tree in the middle. This tree
differs
from the BME tree significantly. A surprising result is that the tree on
the left is almost {\em never} the NJ tree. We believe that a deeper
understanding of the
``mistakes'' NJ makes when it does not optimize the balanced minimum
evolution
criterion may be important in interpreting the results, especially for
large trees. 

We also computed analogous results for $n=6,7,8,9,10$. They 
are available, together with the software for computing volumes, at
the supplementary materials website

\begin{center}
{\tt http://bio.math.berkeley.edu/NJBME}
\end{center}

\section{Discussion}

Theoretical studies of the neighbor-joining algorithm have focused on
statistical
consistency and the robustness of the algorithm to small perturbations
of tree metrics. The paper by \cite{Studier1988} established the
consistency of NJ,
that is, if $D_T$ is a tree metric then NJ outputs the tree $T$. This
result was then extended in \cite{Atteson99} and more recently
\cite{Mihaescu2007} who show that if $D$ is ``close'' to a tree metric
$D_T$
for some $T$, then NJ outputs $T$ on input $D$. 

Our results provide a different perspective on the NJ
algorithm. Namely, we address the question of the accuracy of the
greedy approach for the underlying linear programming problem of BME
optimization. This led
us to the study of BME polytopes, and the combinatorics of these
polytopes is interesting in its own right:

\begin{question}
Is there a combinatorial criterion for two tree topologies forming an 
edge in the BME polytope, similar to pruning/re-grafting or some other
operation on trees? If so, this could be used to define a combinatorial 
pivoting rule on tree space that could be used in hill-climbing
algorithms 
for phylogenetic reconstruction.  Such a pivoting rule would have 
the advantage that it would be equivalent to performing an edge-walk 
on the BME polytope.  Edge-walking methods are known to 
perform well in practice for solving linear programs. See \cite{fastme}
for an example of a local search approach to finding minimum evolution trees.
\end{question}

Similarly, a better understanding of the combinatorics of the NJ cones
will lead to a clearer view of the strengths and weaknesses of the
neighbor-joining algorithm. A basic problem is the following:

\begin{question}
Find a combinatorial description of the NJ cones for general $n$.  How
many facets/rays are there?
\end{question}

Our computational results lend new insights into the performance of the
NJ and BME algorithms for small trees.  
We have measured the relative sizes of cones for different shapes of
trees, and measured 
the frequencies of the all combinatorial types of discrepancies between
BME and NJ
trees.  In particular, we have observed that the NJ algorithm is least
likely to reproduce the BME tree
when the BME tree is the caterpillar tree.

\begin{conjecture}
The caterpillar tree yields the smallest ratio of spherical cone volumes
vol(NJ $\cap$ BME) / vol(BME) where NJ is the spherical cone volume of a
union of the NJ cones and BME is the spherical cone volume of the BME
cone
for a fixed tree.  In other words, the caterpillar tree is 
the most difficult BME tree topology for the NJ algorithm to reproduce.
\end{conjecture}

Another problem we believe is very important is to extend the results
shown in Figure 4 to large trees. In other words, to understand how
neighbor-joining can fail when it does not succeed in finding the
balanced minimum evolution tree.

\begin{question}
What tree topologies is neighbor-joining likely to pick when it fails to
construct the balanced minimum evolution tree?
\end{question}

There are many other interesting cones related to
distance-based methods that can be considered in this context. For
example, in \cite{Mihaescu2007}, it is shown that the {\em quartet
  consistency} condition is sufficient for neighbor-joining to
reconstruct a tree from a dissimilarity map for $n \leq 7$ leaves. 
  The quartet consistency 
conditions define polyhedral cones (consistency cones) in
$\R^{n \choose 2}$ (see \citep{Mihaescu2007} for details). 
For $n = 4$ taxa the consistency cones cover all of $\R^{4 \choose
  2}$ showing that quartet consistency explains the behavior of
neighbor-joining for all dissimilarity maps. 
Using the second method outlined in Section 4 we succeeded in computing
the volumes of
the consistency cones intersected with the first 
orthant of the sphere for $n = 5$ taxa.  There are 15 cones, all
equivalent under orthogonal transformation, and their union covers
27.93\% of $\R_{+}^{5 \choose 2}$, measured with respect to
spherical volume. In other words, quartet consistency explains the
behavior of neighbor-joining on almost $\frac{1}{3}$ of dissimilarity
maps. 

Such computations are pushing the boundary of computational polyhedral
geometry. For $n \geq 6$ taxa, triangulating a consistency cone is too
unwieldy, although we are confident that
spherical volumes could still be computed using polynomial time
hit-and-run sampling methods for 
volume approximation \cite{Deshpande2006}. Such methods are
complicated and not yet implemented.

Finally, we comment on the example in Section 3 that shows how to
different alignments may lead to the same neighbor-joining tree, whereas
the neighbor-joining tree constructed from a concatenation of
the alignments is different. This result has significant implications for
studies where species trees are constructed from multiple gene families
by combining the data.
\pagebreak
\bibliographystyle{plain}

\end{document}